\documentclass[aps,prb,twocolumn,superscriptaddress,showpacs]{revtex4}
\usepackage{epsfig}
\begin{document}

\title{Anisotropic Superparamagnetism of Monodispersive Cobalt-Platinum Nanocrystals}

\author{F. Wiekhorst}
\affiliation{Institut f\"ur Angewandte Physik und Zentrum f\"ur Mikrostrukturforschung, Universit\"at Hamburg, Jungiusstrasse 11, D-20355 Hamburg, Germany}
\author{E. Shevchenko}
\author{H. Weller}
\affiliation{Institut f\"ur Physikalische Chemie, Universit\"at Hamburg, Bundesstrasse 45, D-20146 Hamburg }
\author{J.~K\"otzler}
\affiliation{Institut f\"ur Angewandte Physik und Zentrum f\"ur Mikrostrukturforschung, Universit\"at Hamburg, Jungiusstrasse 11, D-20355 Hamburg, Germany}
\email[]{koetzler@physnet.uni-hamburg.de}

\date{April 17, 2003}

\begin{abstract}\vspace*{1cm}
Based on the high-temperature organometallic route (Sun \textit{et~al.} Science \textbf{287}, 1989 (2000)), we have synthesized powders containing CoPt$_{3}$ single crystals with mean diameters of 3.3(2)~nm and 6.0(2)~nm and small log-normal widths $\sigma$=0.15(1). In the entire temperature range from 5~K to 400~K, the zero-field cooled susceptibility $\chi(T)$ displays significant deviations from ideal superparamagnetism. Approaching the Curie temperature of 450(10)~K, the deviations arise from the (mean-field) type reduction of the ferromagnetic moments, while below the blocking  temperature $T_{b}, \chi(T)$ is suppressed by the presence of energy barriers, the distributions of which scale with the particle volumes obtained from transmission electron microscopy (TEM). This indication for volume anisotropy is supported by scaling analyses of the shape of the magnetic absorption $\chi''(T,\omega)$ which reveal distribution functions for the barriers being also consistent with the volume distributions observed by TEM. Above 200~K, the magnetization isotherms M(H,T) display Langevin behavior providing 2.5(1)~$\mu_{B}$ per CoPt$_{3}$  in agreement with reports on bulk and thin film CoPt$_{3}$. The non-Langevin shape of the magnetization curves at lower temperatures is for the first time interpreted as \textit{anisotropic} superparamagnetism by taking into account an anisotropy energy of the nanoparticles $E_{A}(T)$. Using the magnitude and temperature variation of $E_A(T)$, the mean energy barriers and 'unphysical' small switching times of the particles obtained from the analyses of $\chi''(T,\omega)$ are explained. Below $T_{b}$ hysteresis loops appear and are quantitatively described by a blocking model, which also ignores particle interactions, but takes the size distributions from TEM and the conventional field dependence of $E_{A}$ into account.
\end{abstract}


\maketitle

\section{INTRODUCTION} 
\begin{figure}
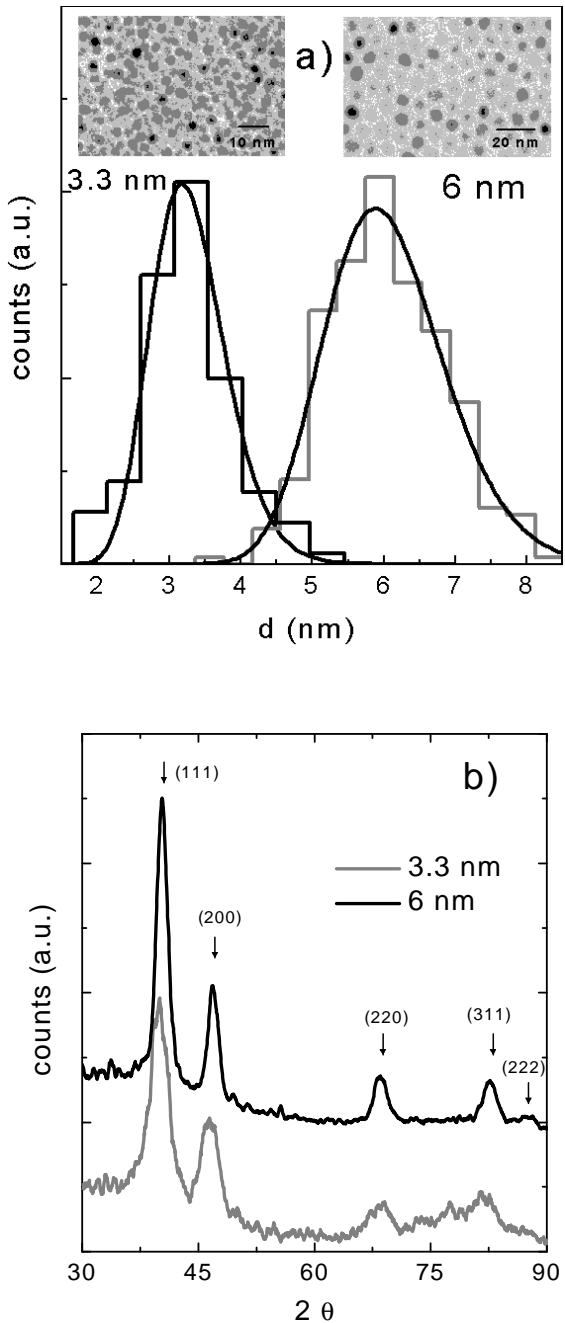

\begin{center}
  \vspace*{2mm}
  \hspace*{-1cm}
	\epsfig{file=fig1a.eps,width=7.5cm}
	\vspace*{-1cm}
	\hspace*{-1cm}\epsfig{file=fig1b.eps,width=9.5cm}
\caption{a) Distribution histograms for particle diameters in CoPt$_{3}$ nanocrystal-powders determined from TEM-pictures, clips of which are displayed by the insets; solid curves represent distribution functions discussed in the text. b) XRD wide-angle scans, which evidence the fcc-structure.}
\end{center}
\end{figure}

 The preparation of ferromagnetic particles suitable for high-density storage media constitutes one of the present challenges to nanotechnology. Most recently, the central demands of this application, i.e.~a narrow size distribution of nanometer crystals and their arrangement in 2- and also 3-dimensional lattices with controllable interparticle spacing, have been met through organometallic synthetic approaches followed by the self-assembly technique. \cite{R1,R2,R2a,R3} The first ferromagnetic nanocrystals prepared by this organometallic route were FePt- \cite{R1} and Co ~\cite{R2} as well, aimed at achieving sufficiently large anisotropy energies at a minimum particle volume $V_{p}$ . This result should drive the thermal fluctuation time $\tau=\tau_{0} \exp (E_{B}/k_{B}T)$ from the microscopic values, $\tau_{0}=10^{-10} -10^{-12}$~s, \cite{R4} beyond the values necessary for the storage stability.\cite{R5} At this point, physical characterization of the nanoparticles is required to explore and understand the origin and the magnitude of the anisotropy constant,$ K_{A}  \approx E_{A}/V_{p}$, which determines the energy barrier $E_{B} \approx  E_{A}$ for coherent rotation of the particle moment $\mu_{p}$. Rather large values of $K_{A} \approx 6 \cdot 10^6$~ J/m$^{3}$ have been achieved for iron-rich Fe$_{x}$Pt$_{1-x}(x \approx$ 0.52 to 0.60) nanoparticles after controlled annealing at high temperatures,\cite{R1} which transformed the fcc to the face-centered tetragonal L1$_{0}$-structure. Due to the larger spin-orbit coupling of cobalt, Co-based nanoparticles may be expected to provide a higher anisotropy, even in the as grown state. In fact, very recently, surprisingly large values of anisotropy up to $2 \cdot 10^6$~J/m$^{3}$ have been reported for 12 nm fcc-Co particles and attributed to the enhancement of $K_{ A}$ at the surface.\cite{R6} In addition, Co$_{\rm x}$Pt$_{\rm 1-x}$ nanoparticles have also been prepared by magnetron sputtering \cite{R7} and a microemulsion technique \cite{R8} with a maximum anisotropy constant $ K_{A} \approx 0.6\cdot 10^6$~J/m$^{3}$ for x=0.75.\cite{R6} Somewhat smaller anisotropy values were obtained for as-grown x=0.25- and annealed x=0.5-particles.\cite{R7} The sources for these anisotropies has not yet be identified, but, considering more detailed studies on annealed Co$_{\rm x}$Pt$_{\rm 1-x}$ films, \cite{R9} internal grain boundaries separating different structures are the most likely candidates for enhanced anisotropies, in addition to surface effects.

In the present work, we present a detailed physical characterization of spherical CoPt$_{3}$-nanocrystals prepared by organometallic route in high boiling coordinating solvents mixtures. \cite{R2a} The possibility to grow 2D and 3D colloidal superstructures using these nanocrystalline spheres, capped by a suitable organic agent to maintain minimum interparticle distances of 2~nm, has been demonstrated in Refs.~\onlinecite{R2a, R3}. Our study is directed towards a determination of the superparamagnetic behavior and the onset of anisotropy in as-grown, single fcc-phase CoPt$_{3}$ nanoparticles. This work is intended to provide a deeper insight into the nature of the magnetic blocking of the single-phase, interaction-free nanocrystals, i.e. in the transition from the Langevin-type superparamagnetism(SPM)- to the blocked  SPM. In the seminal work by Bean and Livingstone,\cite{R10} this dynamical crossover has been defined to occur at the so-called blocking temperature, $T_{b} \cong  E_{B}/25~k_B$, where remanent magnetizations and coercivity appear. For the first time, in this work we also examine the effects of $E_{A}$ on the low-field \textit{equilibrium} magnetization M(H,T), i.e. at temperatures distinctly above $T_{b}$. To this end, we apply a recent general framework of Garcia-Palacios \cite{R11} and take into account the anisotropy in the statistical evaluation of the magnetization for particle assemblies with randomly-distributed anisotropy axes. It turns out that, starting from the isotropic behavior at zero magnetic field H, the magnetization isotherms M(H,T) fall progressively below the commonly-supposed Langevin function ${\cal L}(\mu_{p}H/k_{B}T)$ due to  the presence of a finite anisotropy field as defined by Bean and Livingstone,\cite{R10}, $H_{A}=2E_{A}/\mu_{p}$. We believe, that for large anisotropies, the evaluation of $E_{A}$ from the 'low-field' isotherms is advantageous from that obtained by the frequently used asymptotic law, $M(H,T)\cong M_{0}[{\cal L}(\mu_pH/k_BT)-\frac{1}{15}(H_{A}/H)^{2}]$ (see e.g.~ Ref.~\onlinecite{R12,R13}), because the validity of the latter expression requires rather high fields, $H \gg H_{A}$, which is difficult to reach for materials with strong anisotropy. Moreover, additional paramagnetic contributions from unavoidable impurity phases in the nanoparticle assemblies  may distort the analysis using the asymptotic law. \cite{R13} We hope that our  results will also provide a basis for a further modification of CoPt$_{3}$-nanocrystals in order to optimize the anisotropy.
 
 The outline of this paper is as follows. In Section II, the structural features of the two nanoparticle assemblies under investigation and the magnetic measurements are described. In Section III, first the results of the temperature-dependent low-field susceptibilities are analyzed to extract the temperature variation of the particle spontaneous magnetization, the blocking temperatures, and the effects of the narrow particle size distributions on the blocking behavior of the zero-field cooled (ZFC) susceptibility $\chi$. Then we present AC-susceptibilities, from which the thermal activation barriers and their distribution functions are determined. These distribution functions are compared with those obtained from $\chi$ and the TEM images as well. In Section IV, we report on magnetization isotherms recorded between 5~K and 350~K. First, from the Langevin behavior observed at high temperatures the mean magnetic moments of the particles $\mu_p$ and per CoPt$_{3}$ are deduced. Then, approaching the blocking temperatures from above, the increasing effect of a temperature dependent anisotropy is observed and evaluated. The extrapolation of the resulting $E_{A}$(T) to low temperatures yields energies consistent with the barriers determined from the AC-susceptibilities and yields an anisotropy energy density of $K_{A}=0.12~\cdot~10^6$~J/m$^{3}$ independent of the nanoparticle volume. Finally, the hysteresis loops in the blocked SPM regime, $ T < T _{b}$, are presented and analyzed based on the particle size distributions and the anisotropic SPM magnetization. Section V closes the work with conclusions.

\section{EXPERIMENTAL}

The organometallic route, by which the present assemblies of size controlled nanocrystals were prepared, has been described in detail in Ref. \onlinecite{R2a}. The Co:Pt=1:3 composition of the nanocrystals was obtained from elemental analysis using inductively coupled plasma-atomic emission spectroscopy \cite{R2a}. Transmission electron microscopy (TEM) images of both particle assemblies and their analyses are depicted in Fig.~1(a). The TEM pictures indicate rather narrow distributions of the particle diameters d, which can be nicely fitted to the frequently observed log-normal function, $P(d)= (\sqrt{2\pi}\sigma_{d}d)^{-1}$ exp$(-\ln ^2 (d/d_m )/2\sigma_d ^2)$.  One finds rather narrow size distribution widths, $\sigma_{d}$=0.16 and $\sigma_{d}$=0.14 and from the peak position of $P(v)$, $d_m$, the mean  particle diameters $d_p=\exp(\sigma_d^2/2)d_m= 3.3$~nm and $6.0$~nm, see Table I below. Wide angle X-ray diffraction (XRD) scans, recorded on the pure nanocrystalline powders using $CuK_{\alpha}$-radiation (Philips X-pert) are shown in Fig.~1(b). They reveal the chemically disordered crystalline fcc-phase with lattice constant $a_{0}=3.86~\AA$, consistent with the bulk value \cite{R14}. The widths of the Bragg-peaks have been shown to agree with the particles sizes as determined from TEM images \cite{R2a}. No indication of the chemically ordered (L1$_2$) phase has been detected. The disordered fcc phase is supported by an enhanced value of the Curie temperature $T_c = 450 K$, which we found from a mean-field based estimate presented in section III.A, to be in good agreement with the report by Sanchez et al. \cite{R14a} and in stark contrast to $T_C=300~K$, as determined independently by Ref.\onlinecite{R14a} and more recently by Kim et al. \cite{R14} for the L1$_2$ phase.

All magnetic measurements, i.~e.~the temperature variation of the low-field magnetizations and also the field sweeps up to 10 kOe at fixed temperatures between 5~K and 400~K have been performed using a SQUID-magnetometer (QUANTUM DESIGN, MPMS 2). By using an AC-option we investigated the dynamic susceptibility, $\chi'-i\chi''$, between 0.1 Hz and 1 kHz at H=0, where the excitation amplitude was kept small enough to detect the linear response. As an optimum (root mean square) sensivity for magnetic moment we reached $10^{-8}$ emu. This allowed to investigate the powder samples of 5~mg weight, and 1.5~mm$^{3}$ volume to a high accuracy. The diamagnetic background of the teflon holder has been determined separately.\\

\begin{table*}
\caption{Parameters of the two nanoparticle assemblies determined from the analyses described in the text.}


\begin{tabular}{|c||c|c|c|c|c|c||c|c|c|c|c||} \hline
sample&d(nm)&$\sigma_{\rm{d}}$& $V_p$ (nm$^{3}$) & $T_{b}$(K) & $E_{m}$(K)&$\tau_0$(s$^{-1}$)& $N_p$ (g$^{-1}$) &$\mu_{p}(0)(\mu_{B})$ & $\mu_{\rm{CoPt_3}}(\mu_{B})$ &  $E_A$(0)(K) &$K_{A}  (\frac{10^6 \rm{J}}{\rm{m}^3}$) \\ \hline  \hline
CoPt$_3$-3   &     3.3   &   0.16  & 18.8 &   8.3   &  178    & $2 \cdot 10^{-13}$ &  4$\cdot 10^{17}$  &  785 &   2.4    & 125 & 0.11   \\ \hline
CoPt$_3$-6   &    6.0    &   0.14  &  131 &   37.5     &   990   &  $ 1.6\cdot 10^{-15}$  &2$\cdot 10^{17}$ & 5120  & 2.5  & 770 &   0.10   \\ 
\hline
\end{tabular}
\end{table*}

\section{ZERO-FIELD SUSCEPTIBILITIES}

\subsection{DC-Limit}

\begin{figure}
\vspace*{5mm}
\begin{center}
	\hspace*{-3mm}\epsfig{file=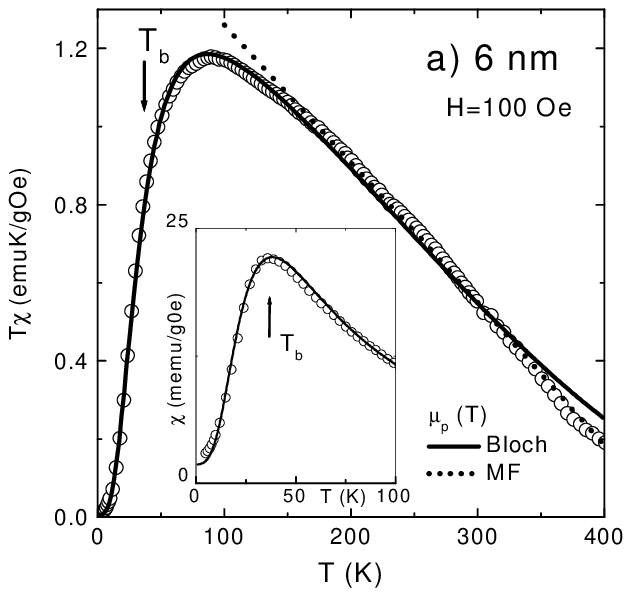,width=8.5cm}
	\vspace*{20mm}
  \hspace*{-3mm}\epsfig{file=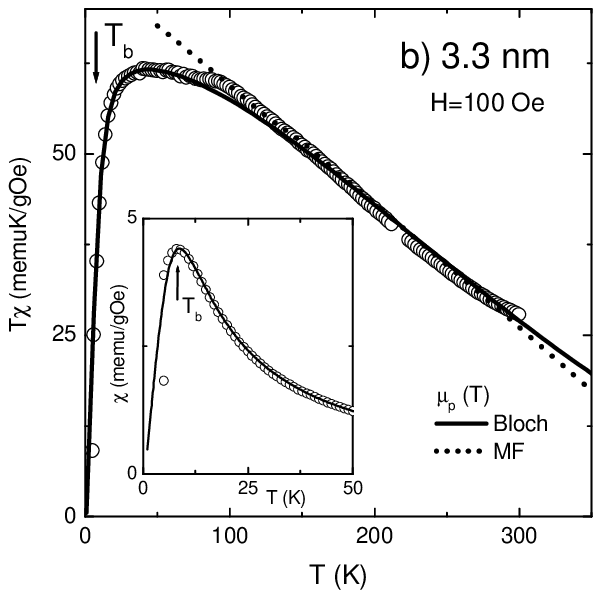,width=8.5cm}
\caption{Temperature dependence of the effective Curie-constants  $\chi T$ determined from the low-field ZFC magnetizations $\chi=M_{ZFC}/H$ (insets) for the nanoparticle powders a) CoPt-6 and b) CoPt-3. Note that for large T, the $\chi$T extrapolate (dashed lines) to the same mean-field Curie-temperature, $T_{MF}=450(20)~K$. The low temperature regime is dominated by a progressive blocking of particles being described by Equ.1 using the log-normal distributions infered from Fig.1b. The solid curves in the insets are calculations from Equ.1 using Bloch's law, $\mu_p(0)(1-B T^{3/2})$ , for the moments at low temperatures.}
\end{center}
\end{figure}

The temperature dependence of the ZFC-susceptibilities of the samples has been determined from the magnetizations measured during warming in a field of 100~Oe. This field was sufficiently weak, $\mu_pH \ll k_BT$, to approximate the zero-field limit, $\chi(T)=M_{ZFC}(T,H)/H$. The only exceptions from $\mu_pH \ll k_BT$ occur at the lowest temperatures for the CoPt$_{3}$-6 sample, but there corrections for finite field can easily be taken into account, in the analysis of $\chi(T,H)$.
The insets to Fig.~2 show that the susceptibilities display clear maxima, which define the blocking temperatures $T_{b}$, see Table I. Determinations of the blocking temperatures themselves allow a first estimate of the mean energy barriers against coherent rotation of the particle magnetic moments $\vec{\mu}_{p}$. Use of the classical estimate of the energy barrier, $E_{B}=\gamma k_{B} T_{b}$, with $\gamma=\ln(t_0/\tau_0)~\cong 25$\cite{R10} and the $T_b$-values listed in Table I yields values of about 950~K and 200~K, which roughly scale with the mean particle volume $V_p$. This result indicates that contributions by surface anisotropy to $E_B$ are small, because they are proportional to $V_p^{2/3}$. 

More detailed insight into the blocking process and also into the magnetism above $T_{b}$ is gained by the effective Curie constants $C_{eff}(T) \equiv \chi \cdot T$, depicted by the main frames of Fig.~2 for both powders. This quantity has been evaluated to show (i) the gradual transition from the Langevin SPM to the blocked SPM when $T_{b}$ is approached from above and (ii) a linear decay of $C_{eff}(T)$ towards higher temperatures in the SPM phase (see dotted lines in Fig.~2). By using the Langevin result for the Curie constant $C_0(T)= \mu^{2}_{p}(T)/3k_{B}\mu_{0}V_{p}\rho$ ($\rho$=mass density) and assuming that near the Curie temperature the spontaneous particle moments display a mean-field (MF) like behavior, $\mu_{p}(T)=\mu_{p}(0)(1-T/T_{c})^{1/2}$, we find as an estimate $T_c \approx 450(10)~K$. This value is consistent with early work reporting $T_c=500~K$ \cite{R14b} for CoPt$_3$ as well as with $T_c=460~K$ determined more recently for disordered fcc CoPt$_3$ \cite{R14a} and also CoPt$_3$ films \cite{R9}. With regard to the evaluation of $T_c$, it may be interesting to note two points: (i) a MF-law for $\mu_{p}(T)$ was realized in recent Monte Carlo simulations by Altbir et al.\cite{R15} for nanosized Co-particles, and (ii) since the data for the 6 nm particles extend until 400~K, the uncertainity for $T_c$ is smaller ($\le$ 10~K) as for the 3 nm particles with T $\le$ 300~K. However, this does not affect our main conclusion in Section II, that the present nanocrystals are chemically disordered, as no indication for $T_c$=300~K of the L1$_2$ phase is observed.

As shown by Fig.2, at lower temperatures the MF-law for $C_0$(T) diplays a rather wide overlap with the temperature variation resulting from
Bloch's law for the particle moment $\mu_{p}(T)= \mu_{p}(0) (1-B T^{3/2})$ being used in previous analyses of $\mu_{p}(T)$ for iron nanocrystals. \cite{R16,R17} For both CoPt$_3$ particle assemblies we obtain for the coefficient
$B=0.6 \cdot 10^{-4} K^{-3/2}$ which turned out to be much larger than the bulk value, 3.0 10$^{-6} K^{-3/2}$. \cite{R17}  An enhancement of the Bloch coefficient for nanoparticles was also found on the $Fe$ nanocrystals \cite{R16,R17} and by Monte-Carlo simulations applied to the Heisenberg-model.\cite{R18}

Approaching the blocking temperature $T_{b}$, one realizes from the graph of $ \chi(T) \cdot T$ that, due to the size distribution of the particles, larger particles remain blocked up to temperatures above $T_{b}$. Following Wohlfarth \cite{R19} and Hansen and M{\o}rup, \cite{R20} we describe the blocking effect on the effectice Curie constant of the ZFC susceptibility by the expression 
\begin{eqnarray}
\chi(T)\cdot T &=& C_0(T)\left [\int^{v_{T}}_{0} dv P(v) v + \int^{\infty}_{v_{T}} dv P(v)\gamma v_{T}\right] \nonumber \\ 
&&+ \chi_{bgd}\cdot T, 
\end{eqnarray}
where $C_0(T)$ represents the Curie constant of the freely fluctuating, i.e. SPM moments $\mu_p$, introduced above. $P(v)$ describes the distribution functions of the normalized particle volumes $v=V/V_{m}$, where $V_{m}=\pi d^3_m/6$ is defined by the maximum of $P(v)$. By using a single, thermal activation volume $v_{T} = V_{T}/V_{m}$, this approach divides the particles into two groups: the first term of Equ.~1 accounts for the free rotation of the unblocked smaller moments, while the second one describes the rotations of the blocked (larger) moments within the energy minima produced by their own anisotropy energy $E_B$. Since this is a rather rough approximation we allow $v_{T}$ to deviate from the traditional value $T/T_{b}$ ~\cite{R20} by introducing $v_{T}=T/T_{0}$ with $T_{0}\neq T_{b}$. The difference between our fitted characteristic temperatures $T_0$ and the blocking temperatures $T_b$ will be dicussed below. Finally, the third term in Equ.~1,  accounts for dia- and paramagnetic background susceptibilities,$\chi_{bgd}(T)=C_p/(T+\theta)+\chi_{dia}$, which are small compared to the SPM susceptibilities and are not of interest here. 

The full curves in Fig.~2 have been obtained from fits to Equ.~1 by assuming the log-normal volume distributions $P(v) = \exp(-\ln^2v/2\sigma^{2}_{v})/\sqrt{2\pi}v$ suggested by the TEM images in Fig.~1(b). The fits are rather sensitive to the $P(v)$-shape as well as to the distribution widths, yielding $\sigma_{v}=0.60$ and 0.52 for the 6~nm and 3.3~nm particles, respectively. These standard deviations are only slightly larger than those obtained from the diameter histograms, $\sigma_{v} = 3 \sigma_{d}$ (see Table I). For both particles sizes, the fitted thermal blocking volumes $v_{T}$ yield $T_{0}=0.63(5)T_{b}$, which implies that Equ.~1 defines temperatures $T_{0}$ significantly below the maximum of $\chi$ at $T_{b}$. This shift can be easily explained by calculating $T_{b}$ from $d\chi/dT=0$. Using Equ.~1, one finds for the ratio $v_{b}=T_{b}/T_{0}$ the equation  
 \[
 v_b  = \int\limits_0^{v_b } {dv} \,P(v)v\,/\,v_b\,P(v_b), 
\] 
which can be solved numerically for $v_{b}$ as a function of $\sigma_v$. Inserting the fitted distribution widths we obtain for $v_b=1.70(3)$ which yield $T_0=0.59(2) T_b$, being very close to the observed values for both particles.

\subsection{AC-susceptibility}

\begin{figure}
\begin{center}
	\epsfig{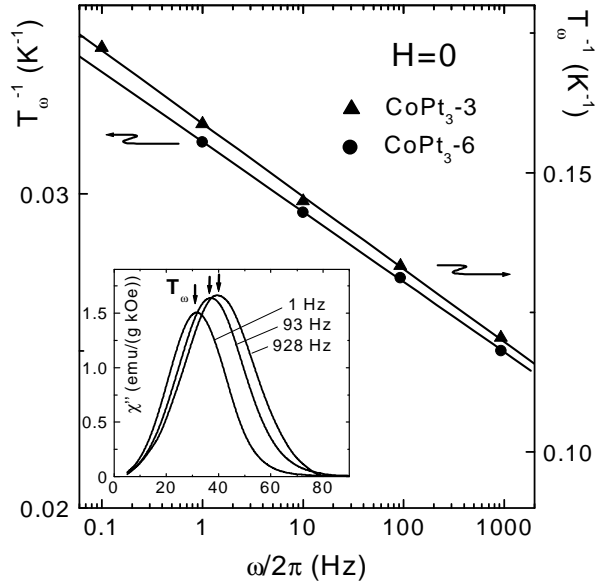}
\caption{Arrhenius plots of the peak temperatures $T_{\omega}$ of the zero-field magnetic absorption curves $\chi''(T,\omega)$, illustrated by the inset. The straight lines are fits providing the mean energy barriers $E_{m}$  and the apparent attempt frequencies $f_{0}=(2\pi \tau_0)^{-1}$ (see Table I).}
\end{center}
\end{figure}

In order to examine the dynamics of the blocking process in some more detail, we have measured the temperature variation of $\chi (\omega,T)$ at fixed frequencies between 0.1 Hz and 1 kHz. Having discussed the contributions to the ZFC susceptibility in Section III.A, we focus here on the portion of susceptibility which relaxes within the measuring period, $2\pi/\omega$, and is observed directly by the loss component $\chi''(\omega,T)$. According to Fig.~3 (inset), $\chi''$ exhibits well-defined maxima at temperatures $T_{\omega}$ that increase with frequency to larger values, as it is typical for a rapid, (Arrhenius-like) relaxation time of the particles, $\tau(T)=\tau_{0} \exp (E_{m}/k_{B}T)$.\cite{R4} The relaxation time at $T_{\omega}$ follows from $\omega\tau(T_{\omega})$=1 and plotting 1/$T_{\omega}$ against $\log( \omega/2\pi)$ in Fig.~3, we obtain straight lines consistent with Arrhenius' law. Note that such analysis only provides a constant activation energy $E_m$, while possible temperature variations of $E_m(T)$ are absorbed by the amplitude $\tau_{0}$. The rather small, apparent switching times of $\tau_0=2\cdot10^{-13}$~s and $2\cdot 10^{-15}$~s are related to $E_m(T)$. We have to postpone the discussion of this feature to Section V. 

As for the blocking temperatures, the results for $E_m$ (Table I) also scale quite nicely with the mean particle volumes $V_m$ and thus indicate the presence of a magnetocrystalline  anisotropy. Relating these barriers to the corresponding blocking temperatures, one finds for the ratios $E_m/k_BT_b=$ 24.9 and 27.5 for the 3.3~nm and 6~nm assemblies, which are very close to the classical estimate of 25.\cite{R10} This widely observed ratio has been estimated by assuming switching times $\tau_{0}=10^{-(11\pm 1)}s$ and measuring times of M$_{ZFC}, t_0=10^{(1\pm1)}s$, which imply $ln~t_0/\tau_{0}=E_{ZFC}/k_{B}T_{b}\approx25$. 

\begin{figure}
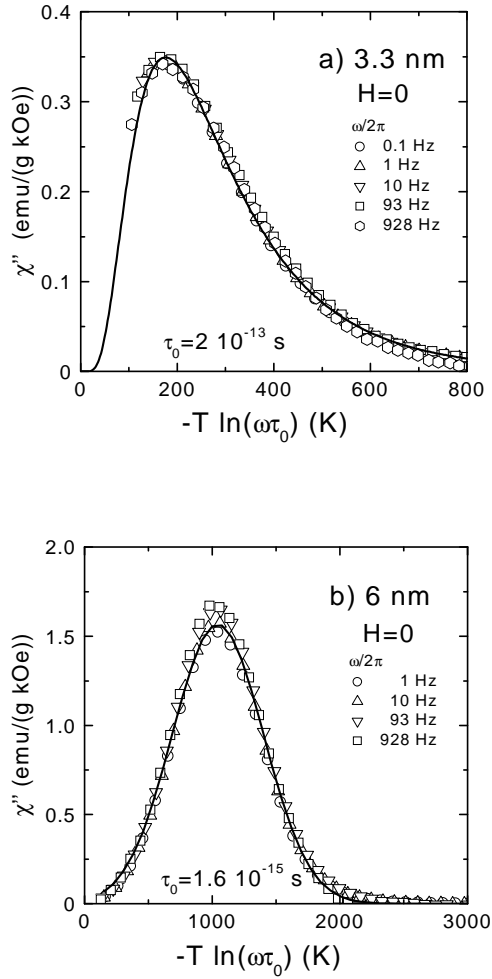

\begin{center}
	\epsfig{file=fig4a.eps,width=7.5cm}
  \epsfig{file=fig4b.eps,width=7.5cm}
\caption{The magnetic absorption of both powders \textit{versus} scaled temperatures; solid curves represent the 'best' distribution functions for the activations energies with peaks at $E_m$, a) log-normal distribution for CoPt-3 and b) gaussian distribution for CoPt-6.}
\end{center}
\end{figure}

For further insight into the dynamics of the particle assemblies, we also analyse the shape of $\chi''(\omega,T)$. We start with a general ansatz, proposed by Shliomis and Stepanov \cite{R21} and applied to experiments by Svedlindh et al.\cite{R22} For non-interacting particles, the anisotropy axes which enclose randomly oriented angles with the probing AC-field, we can write the ansatz as,\cite{R22}
\begin{eqnarray}
\chi (T,\omega )& =& 
\frac{{C_0(T)}}{T}\int\limits_0^\infty  {d\epsilon} \,P(\epsilon)\epsilon
\left\{ {\frac{{R'/R}}{{1 + i\omega \tau (\epsilon)}} + \frac{{1 - R'/R}}{{1 + i\omega \tau_{\bot} }}} \right\}\nonumber \\
 &&+ \,\chi _{bgd}(T). 
\end{eqnarray}

Analogous to the expression for the ZFC-susceptibility, Equ.~1, $\chi(T,\omega)$ consists of a longitudinal and a transverse part describing the inter- and intravalley dynamics of the particles respectively. The relative weights of both contributions are determined by the actual anisotropy $E=\sigma~\cdot~k_{b}T$ of a particle via the statistical factors $R(\sigma ) = \int\limits_0^1 \, dz \,\exp \,(\sigma z^2 )$ and $R'=dR/d\sigma$. The distribution of the barriers against a coherent rotation of the particle moments $\vec{\mu}_p$, $E = \epsilon E_{m}$, is described by $P(\epsilon)$. The small background term $\chi_{bgd}$ proved to be real, i.e. frequency independent, and does not contribute to the imaginary part of $\chi(T,\omega)$. Since the relaxation time of the (transverse) intravalley motion $\tau_{\bot} \approx \tau_{0}$ is much shorter than the (longitudinal) overbarrier time of the nanoparticles, $\tau (\epsilon) = \tau_0 \,\exp \,(\epsilon E_m /k_B T)$, the second term can be ignored in the absorption for the present range of frequencies. Moreover, $\tau(\epsilon)$ varies rapidly as compared to $\epsilon P(\epsilon)$ , so one can safely substitute under the integral for $\chi''(\omega,T)$ in Equ.~2, $\omega \tau /(1 + (\omega \tau )^2 )\approx \,\frac{\pi}{2}\,k_B T\, \cdot \,\delta (\epsilon - \epsilon{_\omega})$,\cite{R11} to yield: 
\begin{equation}
 \chi ''(\omega ,T)\, = \,\frac{\pi }{2}\,\frac{{k_B C(T)}}{{E_B }}\,\,\frac{{R'(\epsilon_\omega  )}}{{R(\epsilon_\omega  )}}\,\,P(\epsilon_\omega  )\,\epsilon_\omega.
\end{equation}

\noindent
Here, $\epsilon_{\omega}=T/T_{\omega}=k_B T(-\ln \omega \tau _0 )/E_m$ designates the maximum relative barrier, over which a particle can thermally jump within the given observation time $2 \pi /\omega$. Therefore, $\epsilon_{\omega}$ is the analogue to $\epsilon_{b}$, with $-\ln \tau_{0}/t_{0}=25$ used before in the discussion of the ZFC susceptibilities. In the present approximation, the absorption $\chi''$ just picks up this 'dynamical' fraction $P(\epsilon_{\omega})d \epsilon$ of the distribution. Except for $P(\epsilon_{\omega})$, the other factors in Equ.~1 vary little as compared to the distribution function. This includes the ratio $R' (\epsilon_{\omega})/R(\epsilon_{\omega})$, which for $E_{m}/k_B T_{\omega}\approx- \ln\omega\tau_{0}>>1$ is always close to one, $R'/R=1-k_B T_{\omega}/E_{m}$.\cite{R11} Hence, in a plot of $\chi''(\omega,T)$ vs. the scaled temperature $-T \ln\omega\tau_{0}=\epsilon_{\omega}E_{m}/k_{B}$ all data should collapse on a single curve. According to Equ.~3, this universal plot provides the distribution functions for the energy barriers. 

The validity of this scaling of $\chi''(T,\omega)$ is demonstrated by Fig.~4 for both nanoparticle assemblies. In the case of CoPt$_{3}$-3 they clearly reveal the same log-normal distribution which already has been obtained from the fit of $T \chi$ in Fig.~2(b). There we found a slightly smaller width of the volume distribution than, $\sigma_{E} \approx 0.6$, for the barriers, which implies for the average barrier $E_B=E_m\exp(\sigma_E^2/2)=195$ K, see Table I. For CoPt$_{3}$-6 a larger difference occurs between the 'volume' distribution functions P(v), as obtained from TEM and $\chi(T)$ on the one hand, and $P(\epsilon)$ from the scaling of $\chi''(\omega,T)$ in Fig.~4(b) on the other hand. The latter unambiguously reveals a gaussian function for the energy distribution with $E_B=E_m$ (Table I). Although one cannot \textit{a priori} expect that the volume and energy distributions agree, the origin for this difference is not clear. It may be interesting to note, however, that very recently the same change, i.e.~from log-normal to gaussian energy distributions, has been detected in magnetic noise spectra, in going from 3~nm to 5~nm Co-particles.\cite{R23} Let us also note, that the amplitudes of the scaled absorptions in Fig.~4 agree \textit{quantitatively} with those predicted by Equ.~3, if the known Curie-constants $C_0$(T) and average barriers $E_{B}$ are inserted. We consider this a confirmation of the validity of the present model.

\section{MAGNETIZATION ISOTHERMS}

\subsection{Isotropic Superparamagnetism} 

\begin{figure}
\begin{center}
  \vspace*{-7mm}
	\epsfig{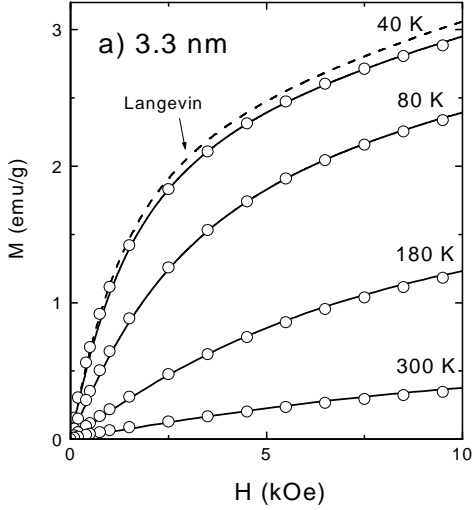}
	\vspace*{-1.5cm}	
  \epsfig{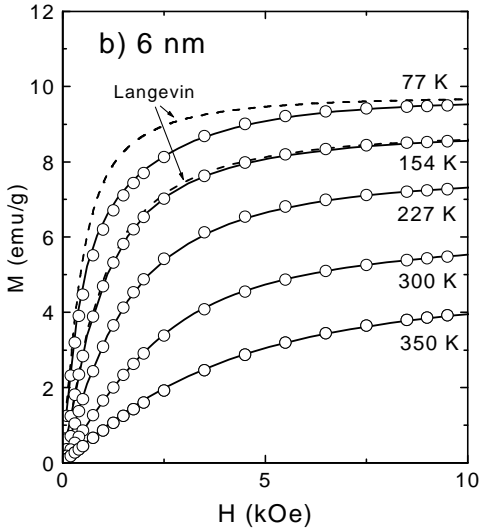}
  \vspace*{1cm}
\caption{Magnetization isotherms recorded above the blocking temperatures $T_{b}$ of the a) CoPt-3 and b) CoPt-6 samples. The solid lines are fits to Equ.7 taking into account a finite anisotropy energy, $E_{A}$(T), describing the significant differences from Langevin-behavior (dotted lines) at low temperatures.}
\end{center}
\end{figure}

The field-dependent magnetization curves $M(H,T)$, recorded above the zero-field blocking temperatures $T_{b}$ of both samples, are anhysteretic, i.~e.~reversible. At temperatures above $T_b$, our main objective is to determine the mean magnetic moment of the nanocrystals $\mu_{p}$ and to investigate the effects of the anisotropy energy $E_A$ on $M(H,T)$. For $E_A\leq k_B T$, the influence of $E_A$ on the magnetization is small, so that the traditional analysis based on the Langevin-function, ${\cal L} (x)=1/$tanh$(x)-1/x$, represents a good approximation to evaluate $\mu_{p}=x \cdot k_{B}T/H$ from the magnetization isotherms using
\begin{equation}
	M(H,T)=N_p \int_0 ^{\infty} dv P(v) \mu_p \, \, {\cal{L}} \left( \frac{v \mu_p H}{k T}\right).
\end{equation}

Our results for $M(H,T)$ are shown in Fig.~5 at selected temperatures above $T_b$. In fact, for temperatures above $\sim$200~K the isotherms can be well described by the Langevin model, i.e. neglecting any anisotropy, if one uses the temperature variation of the moments $\mu_{p}(T)$ explored using the susceptibility (Fig.2). The most interesting quantities emerging from these fits are the maximum particle moments $\mu_{p}(0)$ and the particle densities $N_p$ being listed in Table I. From $\mu_p(0)$ we determine the moments per CoPt$_{3}$ unit using the volume of $57 \AA^3$ per CoPt$_3$ in the fcc-structure. We find $2.4~\mu_{B}$ and $2.5~\mu_{B}$ in the 3.3~nm- and 6~nm-particles being rather close to each other, so that surface effects seem to play no role. These moments per CoPt$_3$ unit are rather close to the bulk values of $2.42~\mu_{B}$ determined by neutron scattering \cite{R24} and $2.6~\mu_{B}$ following from the saturation magnetization measured in fields up to 330~kOe.\cite{R14} All these results turn out to be smaller than the value of $\approx 2.73~\mu_{B}$ \cite{R25} obtained from band structure calculations for CoPt$_{3}$, which predict $1.86~\mu_{B}$ for Co and $0.29~\mu_{B}$ for each Pt. Such high moments have been reported for CoPt$_{3}$ films \cite{R9} grown at some elevated temperature, $T_{s}=400^{\circ}$~C, which also produced a strongly enhanced anisotropy, $0.6\cdot 10^6$~J/m$^{3}$ at 300~K. At lower deposition temperatures, $T_{s} \leq 200~K$, the moments of the films decreased to $2.2~\mu_{B}$, while the anisotropy vanished above 300~K. These remarkable effects were related to the formation of fine Co-platelets in the films. 

The mean particle density $N_p$ obtained for both assemblies and the measured bulk density $\rho =3.5$ g/cm$^{-3}$ can be used to evaluate a mean distance between the particles D$_{nn}\approx (N_p \cdot \rho)^{-1/3}$ and, hence, the effect of their dipole-dipole interaction. The strongest effect is expected for the 6~nm particles, where we find D$_{nn}\cong$ 12~nm and $E_{dd}=\mu_p^2/4\pi\mu_0 D_{nn}^3=8.5~k_B K$, while for the 3.3~nm particles $E_{dd}/k_B=0.5~K$, turns out to be negligible at all temperatures of interest here, T $\geq$ 5~K. These features justify proceeding the analysis of the magnetization curves toward our lowest temperatures of 5~K by using the \textit{pure} interaction-free models.

\subsection{Anisotropic Superparamagnetism} 

\begin{figure}
\begin{center}
	\epsfig{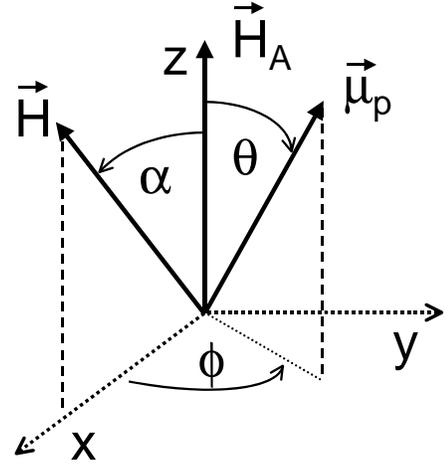}
\caption{Definitions of the angles used in the calculation of the magnetization in the anisotropic superparamagnetic regime.}
\end{center}
\end{figure}

Upon decreasing the temperature but still above $T_b$, the magnetization isotherms begin to fall below the Langevin-curves, an effect we now attribute to the onset of anisotropy. In order to facilitate the computations of $M(H,T)$, we assume the existence of an uniaxial anisotropy, as it is done in most of the literature discussing the dynamical crossover at $T_b$. For randomly distributed axes the influence of anisotropy appears only in finite fields while in zero field the anisotropy effects cancel.\cite{R11}

\begin{figure}
\begin{center}
	\epsfig{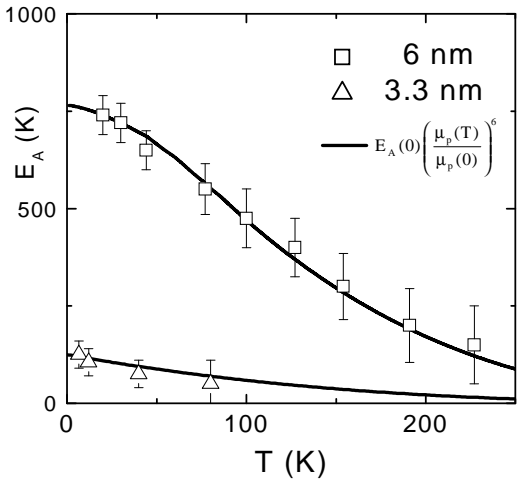}
\caption{Temperature variations of the anisotropy energies resulting from the fits to the reversible magnetization isotherms in Fig.5. The solid curves describe the temperature variations of $E_A(T)$ in terms of the sixth power of the  (spontaneous) particle moments, $\mu_p(T)$ .}
\end{center}
\end{figure}

We start with the Hamiltonian of a anisotropic nanoparticle moment $\overrightarrow \mu  _p$ in a magnetic field $ \vec{H}$,

	\[
{\cal H}= \, - \,E_A \cos^2 \,\theta\, - \,\overrightarrow \mu  _p  \cdot \,\overrightarrow H, 
\]
where $\vec\mu_p$ encloses the angle $\theta$ with the easy axis. Following Garcia-Palacios \cite{R11}, we use the coordinate system displayed in Fig.~6 to calculate the partition function of the particle with the volume $V=vV_{m}$:

\begin{eqnarray}
Z(H,T,\alpha,v)&=&\int\limits_{-\pi}^0 d\,(\cos \theta) \times \nonumber \\
 &&\exp \left[\frac{v(E_A\cos^2 \theta  + \mu _m H\cos \alpha \cos \theta)}{k_BT}\right] \times \nonumber\\
  &&I_0 \left(\frac{v\mu_m H\,\sin \alpha \sin \theta} {k_BT}\right).
\end{eqnarray}
The last factor, $I_o (y) = \pi^{-1} \int\limits_0^\pi  {dt\,\exp \,(y\,\cos t)}$, represents the modified Bessel-function to order zero, resulting from the integration over the spherical coordinate $\phi$. The magnetization of $N_{p}$ particles per gram with a random orientation $\alpha$ with respect to $\vec{H}$ of the easy axes (in principle, other distributions may be included, but are unlikely here) is calculated from standard thermodynamics. After integration over $\alpha$, we obtain:
\begin{equation}
M(H,T,v) = N_p k_B T\,\frac{1}{2}\,\int\limits_0^\pi  {d\,(\cos \alpha )\, \cdot \,\frac{\partial \ln Z} {\partial H}}.
\end{equation}
Finally, we use the log-normal distributions for the particle volumes, obtained in Section III from the blocking behavior of the ZFC-susceptibilities, to calculate the magnetization of the present particle assemblies:

\begin{equation}
M(H,T)=\int \limits_0^\infty{dv P(v) M(H,T,v)} .
\end{equation}
Although these calculations are somewhat time-consuming, depending on the resolution to which the volume-averaging is carried out, their comparison with the data is straightforward. This is due to the fact that the  temperature variation of the particle moments is known, so that the anisotropy energy $E_{A}$ is the only parameter to be fitted.

In Fig.~5, the influence of $E_{A}$ on both assemblies is shown to become significant at the lowest temperatures. This is demonstrated by a reduction of M(H,T) to below the isotropic (Langevin) limits indicated by dotted curves. Due to the lower $E_{A}$-values of the 3~nm particles, the effect is smaller there and becomes even weaker at higher temperatures. The physical reason for this reduction is traced to the fact that, under the influence of the increasing field, the states with transverse magnetization gain a larger statistical weight. Thus, even for a random distribution of the easy axes, M(H,T) becomes smaller in comparison to the isotropic (Langevin) case. For a special set of parameters, this effect has been shown by a recent calculation.\cite{R11}

Although with increasing temperature, thermal fluctuations tend to drive the magnetization towards Langevin behavior, it is possible to extract $E_A(T)$ from our fits of the equlibrium magnetizations to Equ.~5. The results for $E_{A}(T)$ are depicted in Fig.~7, showing that the anisotropy itself decreases with temperature.  Like the energy barriers $E_B$, determined from the dynamic behavior in Section III, the anisotropy scales with the mean particle volume $V_{p}$, and may therefore also be associated with the bulk CoPt$_3$ phase. In order to discuss the temperature variation of the anisotropy, we relate it to the particle magnetization by the conventional power law, $E_A(T)=E_A(0)(\mu_p(T)/\mu_p(0))^n$. The corresponding best fits yield n$\sim$6 and are indicated in Fig.~7. We do not know of any theoretical predictions for the temperature variation of $E_{A}$ in nanoparticles, to which this result can be compared. As a remarkable feature, however, we should mention, that the amplitudes $E_{A}(0) \approx $ 145~K and 800~K are close to the energy barriers $E_{B}$ determined in  Section III at low temperatures from the blocking and the finite dissipation $\sim\chi''$. Using the mean particle volumes, we find a mean density for the anisotropy energy of $0.10(2)\cdot 10^{6}~$J/m$^{3}$. Whether this value can be enhanced by annealing and a possible generation of Co-rich platelets as in Ref.~\onlinecite{R9} remains a challenge for the future preparation.

\subsection{Blocked Superparamagnetism}

\begin{figure}
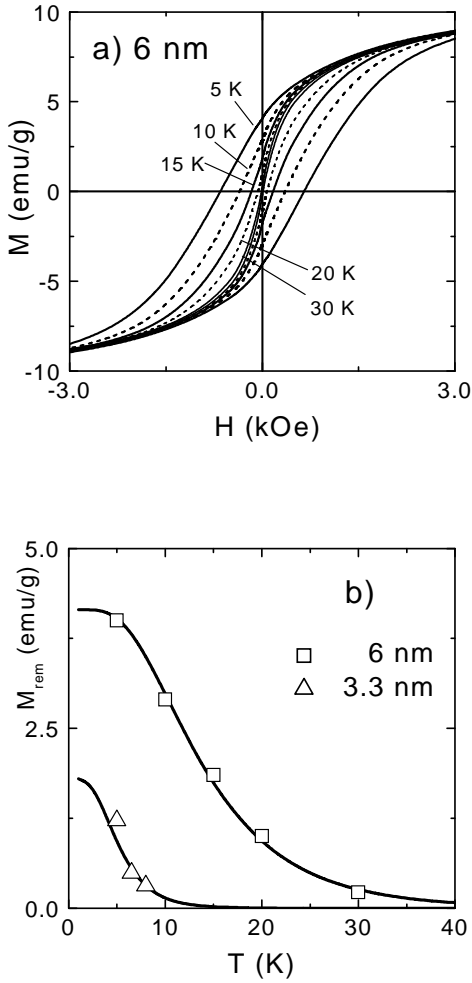

\vspace*{-7mm}
\begin{center}
	\epsfig{file=fig8a.eps,width=7.5cm}
	\epsfig{file=fig8b.eps,width=7.5cm}
\caption{a) Hysteresis loops measured below the blocking temperature $T_{b}$ of 6~nm CoPt$_{3}$-nanoparticles. b) Temperature variation of the remanent magnetizations, the solid curves are fits to Equ.8.}
\end{center}
\end{figure}

We now enter the temperature regime below $T_{b}$ which is characterized by the appearance of hysteresis in the magnetization isotherms, as illustrated by Fig.~8(a) for the 6~nm particles. Except for the lowest temperatures of 5 K we can discuss all results without taking particle-particle interactions into account. This blocked SPM behavior is in contrast to the interacting case, where below some collective ordering temperature spin-glass or - at larger particle densities - long-range ferromagnetism may appear.

\begin{figure}
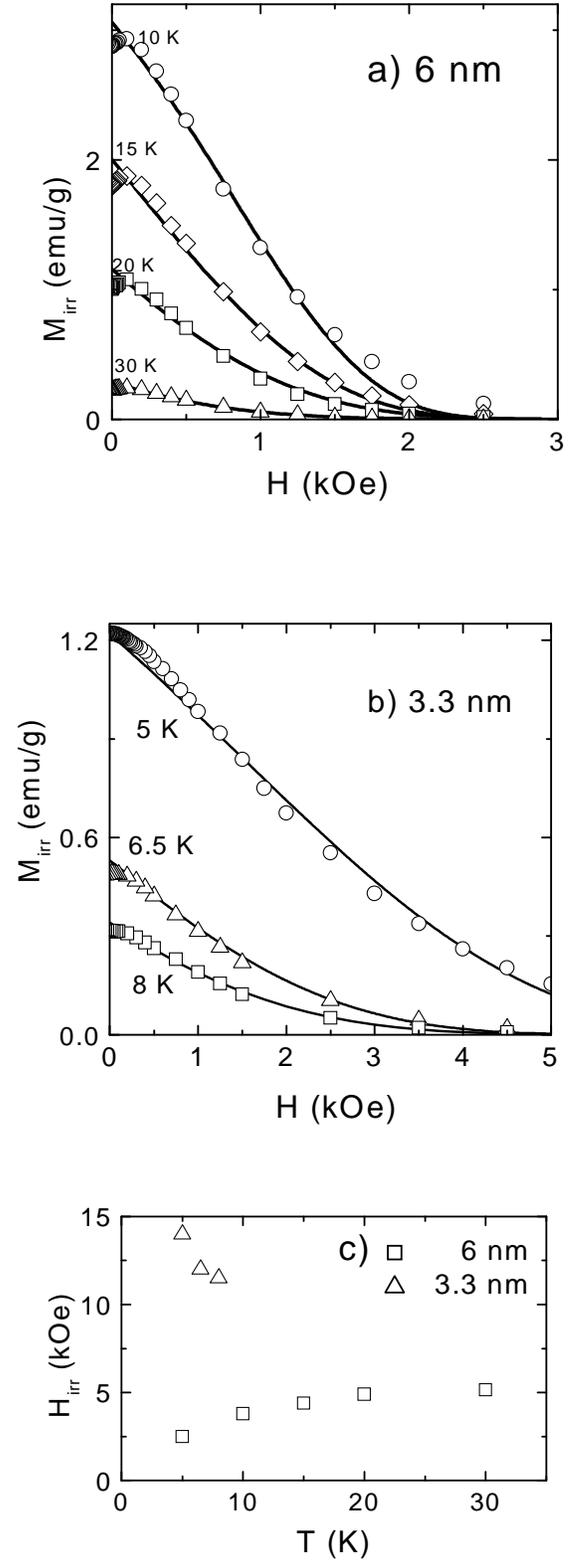

\begin{center}
  \vspace*{-7mm}
	\hspace*{5mm}\epsfig{file=fig9a.eps,width=8.5cm}
	\hspace*{2mm}\epsfig{file=fig9b.eps,width=8.7cm}
	\epsfig{file=fig9c.eps,width=7.5cm}
\caption{Irreversible contributions to the hysteris loops of a) 6~nm (see Fig.8a) and b) 3.3~nm particles. Solid lines are fits according to Equ.9 ; c) temperature variation of the irreversibility fields following from the fits in a) and b).}
\end{center}
\end{figure}

Let us start with the remanent magnetizations measured after sweeps to a maximum field of 10~kOe and shown in Fig.~8(b) for both samples. Within the blocked SPM model of independent particles the most obvious ansatz to describe the temperature variation is 
\begin{equation}
 M_{rem} (T) = M_{rem} (0)\,\,\int\limits_{v_T^*}^\infty  {dv\,P(v)}.
\end{equation}
This form ascribes the remanence to originate from particles larger than a thermal activation volume $v^{*}_{T}$. First, we allowed $v^{*}_{T} = T/T^*_0$ to be different from $v_{T} = T/T_0$ introduced in Equ.~1 to describe the blocking of the ZFC susceptibility. However, as a matter of fact, the best value to fit the data of the 3~nm particles in Fig.~8(b) is $v^{*}_{T}=0.95v_{T}$, i.e. agrees well with $v_{T}$ derived from the SPM susceptibility, Equ.~1. For the 6~nm particles we obtain a larger effective $v^{*}_{T}=1.5v_{T}$. This result implies that the thermal blocking volume of the remanent magnetization is a factor of 1.5 larger than $v_{T}$ obtained from the ZFC susceptibility peak, Equ.~1. One could conjecture that the onset of dipolar interactions between the 6~nm-particles may be responsible for this enhancement of $v^{*}_{T}$. However, the amplitudes resulting from the fits to Equ.~6, i.~e.~$M_{rem}(0)$ =1.9~emu/g and 4.7~emu/g are rather close to 0.5~M(0) (s.Table I), which is fully consistent with the Stoner-Wohlfarth result for interaction-free particles with randomly distributed uniaxial anisotropy axes.\cite{R27} 

As a further extension of this model, we discuss now the field variations of the magnetization below $T_{b}$. To this end, we consider separately the irreversible and reversible contributions, $M_{irr}=(M_{+}-M_{-})/2$ and $M_{rev}=(M_{+}+M_{-})/2$, respectively, where $M_{+}$ and $M_{-}$ denote the branches of the hysteresis loops recorded upon ascending and descending magnetic field. The field dependences of $M_{irr}$ are shown in Fig.~9(a) and 9(b) for both particle assemblies. In the spirit of the analysis of the remanence by Equ.~7, we relate the irreversible magnetization to those particles which still remain to be blocked in the presence of a magnetic field H:
\begin{equation}
M_{irr} (H,T) = M_{rem} (0)\,\int\limits_{v_T^{*}(H)}^\infty P(v)\,dv\quad \quad H < H_{irr}(T).
\end{equation} 
Here $v^{*}_{T}(H)=v^{*}_{T}/(1 - H/H_{irr}(T))^{\beta}$ represents the minimum relative blocking volume which becomes large upon approaching the irreversibility field, where $M_{irr}(H_{irr},T)=0$. This implies that the characteristic field $H_{irr}$ marks the onset of irreversibility in the hysteresis loops. As the best 'simple' exponent to describe the field variation of $v^{*}_{T}(H)$ we found $\beta=2$, which was introduced by Bean and Livingstone \cite{R10} for the field dependence of the particle anisotropy energy. This exponent produces rather nice fits to Equ.~9 (see Fig.~9(a) and 9(b)) using the amplitude from Equ.~8 so that the effective irreversibility field $H_{irr}$ is the only free parameter. We should mention that only at the lowest temperature, 5~K, $M_{irr}$ of the 6~nm particles  could not be fitted by Equ.~8. Referring to our estimate of the particle interactions in Section IV.A, $T_{dd}=8.5~K$, we may attribute this feature to the onset of dipole-dipole interactions. 

The results for the irreversibility fields are displayed in Fig.~9(c). As the most interesting feature we regard the fact, that for the 6~nm nanoparticles $H_{irr}(T)$ agrees almost perfectly with the anisotropy field resulting from the low temperature anisotropy energy. Using the values of Table I, we obtain $H_{A}=2E_{A}/\mu_{p}=4.7~$kOe . For the 3~nm particles the data of Table I yield the same anisotropy field, which qualifies this quantity together with $K_{A}$ , as a bulk property. For $d_p$=3~nm, however, the irreversibility field of $H_{irr}=13(1)~$kOe turns out to be much larger than $H_{A}$. We tentatively attribute this feature to the much larger paramagnetic background in this sample, $\chi_{p}(T)=C_{p}/(T+50K)$. Associating the Curie-constant $C_{p}$ with paramagnetic moment with moments $\mu \approx \mu_B$, we find a fraction of $\approx 30~\%$ of this phase. We conjecture that at the low temperatures of interest here the paramagnetic moments are polarized in the local fields of the oriented nanocrystals so that the effective blocking volume and, hence, the irreversibility field are enhanced.

Finally we apply the present model to the reversible magnetizations. In Fig.~10 is shown just one magnetization isotherm at a low temperature for each sample, where the blocking effects are largest. In order to describe the data, we now assume that only unblocked particles contribute to $M_{rev}$. These particles have volumes smaller than $v^{*}_{T}(H)$ and provide the anisotropic SPM magnetization which can be calculated from Equ.~5,
\begin{equation}
M_{rev} (H,T)=\int\limits_0^{v_T^* (H)} {dv\,P(v)\,M(H,T,v )\,}.
\end{equation}
The results are also indicated in Fig.~10 and show excellent agreement with the data for both nanocrystalline assemblies. Since the same is true for all larger temperatures, we have achieved here a complete description of the hysteretic magnetizations.

\begin{figure}
\vspace*{-1cm}
\begin{center}
	\epsfig{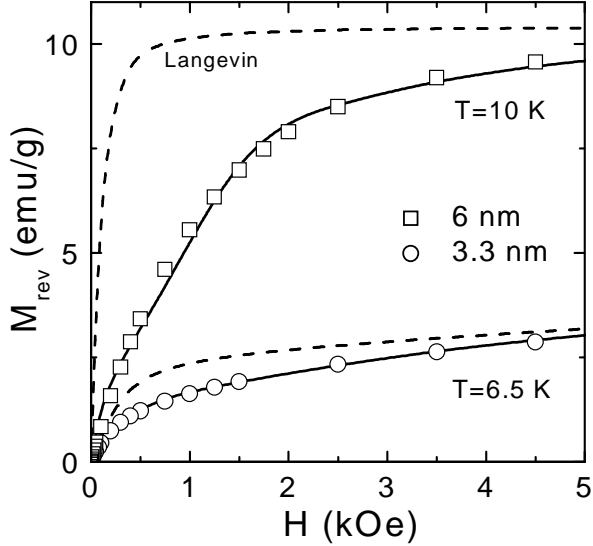}\vspace*{-7mm}
\caption{Reversible part of the hysteresis loops for two low temperatures of both nanoparticle assemblies. Solid lines represent \textit{ab initio} calculations from Equ.10, using the same maximum unblocked volume $v_T^*(H)$ as determined from the fits of the irreversible magnetizations in Fig.9. For comparison, the isotropic Langevin-functions for both samples are indicated.}
\end{center}
\end{figure}

\section{Conclusions}

Our investigations of the zero-field DC- and AC- susceptibilities and field dependent magnetizations of two  CoPt$_3$-nanoparticle assemblies with mean diameters of 3.3~nm and 6~nm provide the first clear evidence for anisotropic superparamagnetism (ASPM). The signature of the ASPM is a \textit{reduced equilibrium}  magnetization. On the temperature axis, ASPM appears between the conventional Langevin-type SPM present at large $T \geq E_A(T)/k_B$ , where thermal fluctuations override the anisotropy, and the so-called blocked SPM occurring below the temperature, $T_b=E_A(T)/25 k_B$,\cite{R10} which represents a non-equilibrium phase depending on the observation time $t_0$. The reduction of M(T,H) in the ASPM regime of nanoparticles with randomly oriented anisotropy axes, appears only in finite magnetic fields $H$ . This effect has recently been predicted \cite{R11} to arise from a slightly preferred statistical weight for particles with perpendicular orientation of their preferred axis relative to $\vec{H}$ .

We have analysed our M(H,T) curves using the full statistical model and deduced a rather strong temperature and size variation of the (uniaxial) anisotropy $E_A(T)$. The linear variation of $E_A$ with the particle volumes reveals the dominance of a bulk anisotropy density of $K_A(0)=0.12 \cdot 10^6$~J/m$^3$. Its temperature dependence could be described in terms of the spontaneous particle moments, $E_A(T)\sim \mu_p^6(T)$, but the origin of this exponent is not yet known. This result implies that anisotropy effects in the present CoPt$_3$ nanoparticles  become important at low temperatures. The temperature variation of $E_A(T)$ should receive also attention for alloys with enhanced anisotropies, as prepared recently for possible high density storage fabrication.\cite{R1,R2,R5,R7} In such materials, of course, the transition to the blocked state may be shifted to beyond room temperatures, but the thermal stability of the blocked state depends on $E_A(T) \sim \mu_p^n(T)$, that is on the exponent $n$ and on the Curie temperature. In the blocking regime, $T < T_b$, we could explain the hysteretic magnetization curves quantitatively within the ASPM model considering blocking and the independently determined volume distribution functions.

Finally, we point out an interesting consequence emerging from the temperature variation of the anisotropy $E_A(T)$. This refers to the activation energy $E_m$ and the time scale $\tau_0$ of the Arrhenius' law which is traditionally used to determine a temperature independent anisotropy constant $K_A$ from blocking phenomena, like the magnetic absorption $\chi''(\omega,T)$ or peaks of $\chi(T)$. Our analyses of $\chi''$ in Section III.~B produced values (i) for $E_m$, which were larger than the anisotropy energies, determined in the ASPM regime (s. Fig.~7), and (ii) for $\tau_0$, which appeared unphysically small and strongly size-dependent (s. Table I). Both features can be understood by starting from the fact that high barriers imply a rather narrow temperature range of $T_{\omega}$ which is available for the Arrhenius' analysis, see inset to Fig.~3. Therefore, to lowest order one may account for the temperature variation of $E_A$ by a linear expansion around some mean temperature $T_{\overline{\omega}}$ from the experimental range: 
	\[
	E_A(T)=E_A(T_{\overline{\omega}})+E_A'(T_{\overline{\omega}})(T-T_{\overline{\omega}})+...
\]
where $E_A'(T_{\overline{\omega}})=(d E_A/dT)_{T=T_{\overline{\omega}}}$. Inserting this as 'true' barrier into Arrhenius' law $\tau_0=\tau_A \exp(E_A(T)/k_B T)$, one finds the same form $\tau =\tau_0 \exp(E_m/k_B T)$, but with renormalized parameters $E_m=E_A(T_{\overline{\omega}})-E_A'(T_{\overline{\omega}})T_{\overline{\omega}}$ and $\tau_0 =\tau_A \exp(E_A'(T_{\overline{\omega}})/k_B)$. Since the anisotropy energy generally decreases with temperature, the conventional analysis overestimates the barrier and produces too small switching times. For the present nanoparticles we found $E_A(T)=E_A(0)(\mu_p(T)/\mu_p(0))^6$, see Fig.~6, where the particle moments obeyed Bloch's law, $\mu_p(T)/\mu_p(0)=1-BT^{3/2}$. Use of this temperature variation oy $\mu_p(T)$ yields for the 'true' barrier against particle switching $E_A(T_{\overline{\omega}})=E_m [1-9(T_{\overline{\omega}}/T_0)^{3/2}]$ and for the real switching time $\tau_A =\tau_0 \exp[(9 E_A(T_{\overline{\omega}})/k_B T_{\overline{\omega}})(T_{\overline{\omega}}/T_0)^{3/2}]$ where $T_0=B^{-2/3}=$640~K was found in Section III.A for the present CoPt$_3$ particles. The strongest effect of $E_A(T)$ on the Arrhenius parameters is expected for the 6~nm particles with $T_{\overline{\omega}}\approx 40~$K, where we obtain $E_A(T_{\overline{\omega}})/k_B =850~$K, close to the results from the magnetization isotherms in the ASPM regime, while for the 3.3~nm particles ($T_{\overline{\omega}}\approx 9~$K) the corrections become negligible.
For the real switching time we obtain $\tau_A=1.0\cdot 10^{-13}~$s, which is close to $\tau_A=\tau_0=2\cdot 10^{-13}~$s for the 3 nm particles. 

The latter results suggest a comparison to the prediction by the N\'{e}el-Brown theory\cite{R4,R11,R26},  $\tau_N=(\pi k_B T/E_A)^{1/2}(\eta+\eta^{-1})/2\gamma H_A$. Since the anisotropy field $H_A=4.8~$kOe and also $\pi k_B T_{\overline{\omega}}/E_A \approx 0.1$ turned out to be independent of the particle size we find for both assemblies $\tau_{N}=(\eta + \eta^{-1})\cdot 1.8\cdot 10^{-12}~$s. Obviously, no value of the Landau-Lifschitz-Gilbert parameter $\eta$ can explain the experimental results for $\tau_A$. As another, rather rough estimate we may assume thermal agitation $\tau_T=\hbar/k_{B}T_\varpi$, which leads to more consistent values of $2 \cdot 10^{-13}~$s and $8 \cdot 10^{-13}~$s for the 6~nm and 3~nm particles, respectively. In order to shed more light into these microscopic dynamics, we presently investigate the ferromagnetic resonance on the CoPt$_3$ nanocrystals.\cite{R28}
 
This work is part of the program of the Graduiertenkolleg 'Physics of Nanostructured Solids' financed by special funds of the Deutsche Forschungsgemeinschaft.

\end{document}